\documentstyle[aps,epsf,twocolumn,prl]{revtex}

\newcommand{\beq}{\begin{equation}} \newcommand{\eeq}{\end{equation}}  
 
\begin{document}

\author{Alexei V. Tkachenko \\
Bell Labs, Lucent Technologies, 600-700 Mountain Ave., Murray Hill, NJ 07974\\
E--mail: alexei@bell-labs.com}
\title{Generalized Entropy approach to far-from-equilibrium statistical mechanics.}
\wideabs{\maketitle
\begin{abstract}
We present a new  approach to far--from--equilibrium statistical mechanics, based on the concept of generalized entropy, which is  a microscopically-defined generalization of  Onsager-Machlup functional. In the case when a set of slow (adiabatic) variables can be chosen, our formalism yields a general form of the macroscopic  evolution law (Generalized Langevin Equation) and extends Fluctuation Dissipative Theorem. It also provides for  a simple understanding of recently--discovered  Fluctuation Theorem.     

\end{abstract}

}
Foundations of non-equilibrium statistical physics remain in focus of intensive research for nearly a century. Over past several decades, there has been a dramatic progress in application of stochastic equations to wide variety of complex systems, as well as in understanding some of their generic features. However, the methods of microscopic derivation of such coarse-grained descriptions (e.g. starting with a Hamiltonian),  remain  essentially limited to traditional kinetic theory and linear response scheme \cite{Groor}\cite{Chai}. In this work, we emphasize the reductionistic mission of non-equilibrium statistical mechanics, by building a constructive formalism whose conceptual framework resembles that of equilibrium theory. Our approach is not limited to the vicinity of thermal equilibrium, and becomes  equivalent to the classical linear response theory in this limit.
     
The central concept in our discussion is {\em Generalized Entropy} (GE), which goes back to works by Onsager and Machlup \cite{Ons}\cite{OnsMach}. Further  development of this paradigm includes its generalization  for far-from-equilibrium case by Graham \cite{Graham} and   action-principle approach to Marcovian  stochastic dynamics by Eyink \cite{Eyink1,Eyink2}, whose technique and conclusions  have many common points with  ours. The distinct feature of the present approach is  an explicit microscopic interpretation of  GE, which enables us to {\em derive} the macroscopic evolution  equations starting from a microscopic level.
Normally, entropy is assigned  to  a  macroscopic {\em  state} of a system as the logarithm of its statistical weight. Similarly, we introduce GE, as a logarithm of the statistical weight of a given macroscopic {\em evolution} (trajectory).
We limit ourselves to the class of   microscopic dynamics (parameterized with microvariables $q_j$, $j=1...N$) which are deterministic and phase-volume-preserving (as e.g. in Hamiltonian systems). In this case, there is a natural measure in the space $q_j$, i.e. the statistical  weight of any   subset of the space  is its  phase volume.   Let   ${\bf A} = {\bf \hat{A}}(q) $ be a  macroscopic (possibly, multicomponent) variable.  The formal definition of the GE associated with a given evolution   ${\bf A}(t)$  between two moments of  time, $t_i$ and $t_f$, is the following: 
\beq
\label{entropy}
S\left[ {\bf A}(t)\right]_{t_i}^{t_f}\equiv \log \left[\int D[q_j]\prod_{t=t_i}^{t_f}\delta\left({\bf \hat{A}}(q(t))-{\bf A}(t)\right)\right].
\eeq
Here, integration is performed over all {\em  actual} microscopic trajectories $q_j(t)$, i.e. those satisfying the microscopic equations of motion. Since we assume the deterministic and phase--volume--preserving  evolution on that level, all the actual trajectories are equally weighted.

Our next step is to distinguish between two contributions to GE: $S\left[ {\bf A}(t)\right]_{t_i}^{t_f}=  S_{\Delta t}^{(0)}(\overline{\bf A})+S^{(kin)} \left[ {\bf \dot{ A}}(t)\right]_{t_i}^{t_f}$.
 The first term, $ S^{(0)}$  is the logarithm of  the total number of trajectories of length $\Delta t\equiv t_f-t_i$ with a fixed "midpoint" ${\bf \overline A}\equiv \left({\bf A}_{t_i}+{\bf A}_{t_f}\right)/2$. The other contribution, $S^{(kin)}$, to which we will refer as {\em kinetic entropy}, is the logarithm of the  probability of   a given  evolution ${\bf \dot{ A}}(t)$ for the fixed $t_i$, $t_f$ and $ \overline {\bf A}$:
\begin{eqnarray}
S^{(kin)}\left[ {\bf \dot{ A}}(t)\right]_{t_i}^{t_f} \equiv 
\log \left\langle \prod_{t=t_i}^{t_f}\delta\left(\frac{d}{dt}{\bf \hat{A}}(q(t))-{\bf \dot A}(t)\right)\right\rangle_{ \overline{\bf A}}=
\nonumber
\\
\log \int D[{\bf X}(t)]\exp\left(\Sigma[{\bf X}(t)]-\int_{t_i}^{t_f} {\bf X}(t)\cdot{\bf \dot A}(t) dt\right).
\end{eqnarray}

Here we have   transformed  the expression by using the exponential representation of $\delta$-function, which resulted in   introduction of a  new variable ${\bf X}$ conjugated to ${\bf \dot A}$. $\Sigma$ is the  generating functional for the variable $\bf \dot A$, which in particularly means that its  variation of any order in $\bf X$ coincides with the corresponding irreducible correlator of the conjugated field (in our case, $\bf \dot A$) :
\begin{eqnarray}
\label{sigma}
\Sigma[{\bf X}(t)]\equiv \log \left\langle \exp \left(\int_{t_i}^{t_f}{\bf X}(t)\cdot\frac{d}{dt}{\bf \hat A}(q_j(t))dt\right)\right\rangle_{ \overline{\bf A}}=
\\
\sum_{n=1}^{\infty}\frac{1}{n!}\int dt_1...dt_n \left\langle\left\langle {\bf \dot A}(t_1)...{\bf \dot A}(t_n)\right\rangle\right\rangle_{ \overline{\bf A}}\cdot {\bf X}(t_1)...{\bf X}(t_n).
\nonumber
\end{eqnarray}

Here $\langle\langle \rangle\rangle$ is the irreducible part of the correlator, i.e. the n-point average  with subtracted contribution  reducible to the lower--order correlation functions. Here and in the future ${\bf \dot A}(t_1){\bf \dot A}(t_2)$ denotes  direct tensor product, which is to be distinguished from scalar one, e.g.  ${\bf X(t)}\cdot{\bf \dot A}(t)$. The averaging is performed over all trajectories with the given initial and final times ($t_i$, $t_f$) and fixed midpoint, $\overline{\bf A}$. 

 Now, after we have related kinetic entropy $S^{kin}$, to the statistics of  $\bf \dot A$, it becomes possible to express  another contribution to GE, $S^{(0)}( \overline {\bf A})$, in terms of regular thermodynamic entropy $S({\bf A})$. For doing so we notice that integration of the statistical weights of all trajectories originating from a given point $\bf A$ is,   by definition, the weight of the initial state, $ \exp S({\bf A})$. After making a simple calculation based on this observation, one gets $S({\bf A})=S^{(0)}({\bf A})+ \Sigma \left[\frac{1}{2}\delta S^{(0)}/\delta \bf A\right]$.
The time interval $\Delta t$, and the associated change of ${\bf A}$ is assumed to  be sufficiently small so that the linear expansion of $S^{(0)}$ in $\Delta {\bf A}$ be valid.

In a general case, $\Sigma$-functional is an   awkward  mathematical object, because of its non--local  structure. However, until this point we have not restricted the choice of the macroparameters, ${\bf  A}$. From the  practical point of view, it is clear that the coarse--grained description of a system may be reasonable  if one can  choose  a set of relatively slow variables as the  macroparameters. Below we specify this choice in a more formal way.

By definition, the form  of $\Sigma$--functional (and the correlators of ${\bf \dot A}$)   depends on the midpoint position, $\overline{\bf A}$. However, there naturally exists an interval $\delta\overline{\bf A}$  within which such dependence can be neglected. This allows one to introduce a concept of drift time, $\tau_{drift}$, over which most of the trajectories remain within this interval of constant statistics of ${\bf \dot A}$. Suppose there exists a shorter  time scale, $\tau_0\ll\tau_{drift}$ such that any correlator $\langle\langle{\bf \dot A}(t_1)...{\bf \dot A}(t_n)\rangle\rangle$ becomes negligible when $\left|t_1-t_n\right|>\tau_0$. In this case,  one can  choose the initial and final times such that $\tau_0\ll\Delta t \equiv t_f-t_i\ll\tau_{drift}$. This considerably simplifies the expression for $\Sigma$--functional: if we are only interested in the behavior of the system on times larger than $\tau_0$, $\Sigma$ becomes local in ${\bf X}$:
\beq
\Sigma[{\bf X}(t)]=\int_{t_i}^{t_f}{\Xi\left({\bf X}(t)\right)dt},
\eeq
here  
\beq
\label{xi}
\Xi\left({\bf X}\right) =  \sum_{n=1}^{\infty}\frac{{\bf X}^n}{n!}\int dt_1... dt_{n-1} \left\langle\left\langle {\bf \dot A}(0)...{\bf \dot A}(t_{n-1})\right\rangle\right\rangle.
\eeq
We will refer to  time scale $\tau_0$  as {\em ergodicity time}. It can be shown that from the methodological point of view, the assumption of the  existence  of such time scale  does  play a role similar   to that of the  ergodic principle in   equilibrium theory.
 
Collecting the above results gives  the following expression for GE:
\beq
\label{hamilton1}
S[{\bf A}(t)]_{t_i}^{t_f}=S\left({\bf A}(t_i)\right)+\log \int D[{\bf X}(t)]\exp S'[{\bf A}(t),{\bf X}(t)],
\eeq
The first contribution here is the conventional entropy of the initial state, and the other one is the logarithm of the probability of a given  evolution starting at that point. The latter is expressed in terms of the functional $S'$, which has a meaning of GE in $({\bf A},{\bf X})$ space:
\beq
\label{hamilton2}
{\cal S'}[{\bf A}(t),{\bf X}(t)]= \int_{t_i}^{t_f}dt\left[ {\bf \dot A}\cdot\left(\frac{1}{2}\frac{\delta S}{\delta {\bf A}}-{\bf X}\right) +\Xi\left({\bf X}\right)- \Xi\left(\frac{1}{2}\frac{\delta S}{\delta {\bf A}}\right)\right].
\eeq
 This quantity becomes additive on the time scales exceeding $\tau_0$, which means that the dynamics becomes Markovian. This fact allows us to extend the applicability of the above expression to the case when $t_f-t_i>\tau_{drift}$, i.e. when  $\Xi({\bf X})$ is no longer  independent of ${\bf A}$. 

In a  particular case of reversible microscopic dynamics, $\Xi$  does not change if the  sign of $\bf X$ reversed. This implies  that  {\em the ratio of probabilities  of   direct and reversed   evolutions along the same path} ${\bf A}(t)$ is independent of the form of $\Sigma$-functional and {\em is given by} $\exp({\bf A}(t_f)-{\bf A}(t_i)$ (see  (\ref{hamilton1})-(\ref{hamilton2})).  This property is known as {\em Fluctuation Theorem} \cite{FT}, which has been recently established for a wide variety of non-equilibrium systems.

One can   eliminate the "fictitious" variable ${\bf X}$ from Eqs. (\ref{hamilton1})-(\ref{hamilton2}):
\begin{eqnarray}
\label{lagrange}
S[{\bf A}(t)]_{t_i}^{t_f}=\frac {S\left({\bf A}(t_i)\right)+S\left({\bf A}(t_f)\right)}{2} +
\\
\int_{t_i}^{t_f}dt\left [\Lambda\left({\bf \dot A}(t)\right)- \Xi\left(\frac{1}{2}\frac{\delta S}{\delta \bf A}\right)\right],
\nonumber
\end{eqnarray}
here the kinetic entropy, which is now   a local functional, has been expressed  as a time  integral of pseudo-Lagrangian $\Lambda$:
\begin{eqnarray}
\label{transform}
S^{(kin)}\left[ {\bf \dot{ A}}(t)\right]=\int_{t_i}^{t_f}\Lambda({\bf \dot A}(t))dt  =
\\ 
 \log \int D[{\bf X}(t)]\exp\int_{t_i}^{t_f}dt \left[\Xi({\bf X}(t))-{\bf X}(t)\cdot{\bf \dot A}(t)\right].
\nonumber
\end{eqnarray}

Because of the obvious analogy with classical mechanics, we will refer the above $({\bf A}, {\bf X})$ and $({\bf A}, {\bf \dot A})$ forms for GE as pseudo--Hamiltonian and pseudo--Lagrangian ones, respectively. Though they are completely equivalent, the pseudo-Hamiltonian formalism requires introduction of additional variables $\bf X$ (which  plays the role of momentum conjugated to measurable $\bf A$), while the pseudo--Lagrangian form  obscures the relationship between the conjugated  functions $\Lambda$ and $\Xi$, given by Eq.(\ref{transform}).   

In the case of a distributed system, when  both ${\bf A}$ and ${\bf X}$ are fields, $\Lambda$ and $\Xi$ would typically become local functionals. If there is a global conservation law for one or several components of ${\bf A}$, the $\Sigma$-functional, Eq. (\ref{sigma}) is invariant with respect to the global transformation ${\bf X}({\vec r})\rightarrow {\bf X}({\vec r})+\sum_{\alpha} \delta^{(\alpha)} {\bf n}^{(\alpha)}$ (here index $\alpha$ counts all the conserved components of  field ${\bf A}$,  ${\bf n}^{(\alpha)}\cdot{\bf A}$). Existence of this transformation means that $\Sigma$ and $\Xi$ should depend only on gradients (and higher spatial derivatives) of the corresponding components of the field  $\bf X$:   $\Sigma=\int \xi(\nabla X^{(\alpha)}) dt d{\vec r} $. If the   microscopic fluxes ${\vec j}^{(\alpha)}$ of the conserved parameters can be introduced, the expansion of $\xi$ in powers of   $\nabla X^{(\alpha)}$ is given by the form similar to Eq. (\ref{xi}), with all the correlators of ${\bf \dot A}$ replaced with those of microfluxes ${\vec j}^{\alpha}$.  

We now discuss the  dynamics of the system in the deterministic limit, which correseponds to the settle point of the GE functional.
One has to emphasize that although  the structure of the functional is similar to regular action, the result of the  variation procedure is dramatically different from that in mechanics. In our case, only  initial point ${\bf A}(t_i)$ should be kept fixed, while the initial rate  ${\bf \dot A}(t_i)$ (or, equivalently, the final point ${\bf  A}(t_f)$), is subject to optimization. As a result, the current value of $ {\bf A}$, rather than the pair $({\bf A},{\bf \dot A})$ determine the future dynamics of the system, and the equation of motion is of the first, rather than the  second order. Namely, maximization of  the pseudo-Lagrangian form of GE, Eq. (\ref{lagrange}), with respect to ${\bf  A}(t_f)$, yields the following dynamic law:
\beq
\label{lagr}
\frac{\delta \Lambda({\bf \dot A})}{\delta {\bf \dot A}}=-\frac{1}{2}\frac{\delta S}{\delta \bf A}.
\eeq
This equation can be interpreted as a balance between thermodynamic driving  force (the right hand side) and the dissipative force (the apparent physical meaning of the left hand side).  
An alternative  description can be obtained by variation of the functional in Eq.(\ref{hamilton2}) with respect to   ${\bf \dot A}$ and $\bf X$:
\begin{eqnarray}
\label{ham}
{\bf \dot A} = \left.  \frac{\delta \Xi}{\delta {\bf X}} \right|_{{\bf X} = \frac{1}{2}{\delta S}/{\delta \bf A}}.
\end{eqnarray}
The equation  shows how the system moves in response to the thermodynamic driving force, which in the  deterministic limit appears to be identical to variable ${\bf X}$. 

In the case of conserved components of $\bf A$-field, Eq. (\ref{ham}), will be  replaced with regular continuity equation, ${\dot A}^{(\alpha)}=-\nabla \cdot \vec{J}^{(\alpha)}$, in which  the macroscopic flux is given by constitutive equation, $\vec{J}^{(\alpha)}=\delta \xi/\delta(\nabla X^{\alpha})$. It is interesting to note that  the macroscopic flux can be introduced even if there is no well-defined fluxes on microscopic scale: this concept   follows from the spatial locality of $\Sigma$-functional and conservation of quantity $\int A^{(\alpha)}d{\vec r}$.

``Kinetic potentials'' $\Lambda$ and $\Xi$  in the deterministic  limit are related through Legander transform, i.e. $\Xi({\bf X})=\Lambda({\bf \dot A})+  {\bf X}\cdot{\bf \dot A}$, where ${\bf X}=-\delta \Lambda({\bf \dot A})/\delta {\bf \dot A}$. In  the vicinity of equilibrium the driving forces are small, and therefore only leading terms  in expansion (\ref{xi}) remain relevant: $\Xi=\langle {\bf \dot A}\rangle\cdot {\bf X}+\frac{1}{2}\Gamma^{(2)}\cdot{\bf X}^2$. After substituting this expression for $\Xi$ into equation of motion (\ref{ham}), we recover the classical linear response result: ${\bf \dot A}=\langle {\bf \dot A}\rangle+\frac{1}{2}{\hat \Gamma}^{(2)}\cdot\delta S/\delta {\bf A}$, i.e. the dissipative contribution to ${\bf \dot A}$ is proportional to the thermodynamic driving force, and our result for the corresponding kinetic coefficient (see (\ref{xi})) coincides with the one given by Fluctuation--Dissipative Theorem (FDT). It is a general practice to assume the  same linear rate-force relationship  even in the regime in which entropy (free energy) is  no longer a harmonic function of deviations from equilibrium. Langevin equation \cite{Chai} is one of the most noticeable examples of such an approach. Furthermore, {\"O}ttinger {\em et al} \cite{Ottinger} have  recently proposed an elegant unified way of representing most of the known stochastic models in a single generic form, which again assumes a linear rate-force relationship for dissipative dynamics. Our Eq. (\ref{ham}) (or (\ref{lagr})) in a general case would  result in a nonlinear relationship between them and, in this sense,  can be referred to as Generalized Langevin Equation (GLE) (an additional noise term will be discussed below).  The fact that a particular form of the evolution equation depends on the correlators of macroparameters, suggests a possibility for a synergy between  our scheme and  earlier   field--theoretical approaches  to non-equilibrium statistical mechanics \cite{msr}.

In order to demonstrate how GLE works outside the linear regime, we consider a trivial  kinetic problem: an ensemble of independent two--state systems, each of which has the same transition rate $\kappa$ in either direction.  The relaxation dynamics for the population difference between the two states, $N_-\equiv N_1-N_2$ is given by equation ${\dot N}_-=-2\kappa N_-$.
Although linear, it is {\em not} a  linear response result. The thermodynamic driving force conjugated to $N_-$ is the chemical potential difference  between the two states, i.e. in our notations
$2X_-= {\partial S}/{\partial N_-}=\log{N_2}/{N_1}$.
This expression can be linearized in $N_-$ (for constant $N_+\equiv N_1+N_2$)
only sufficiently close to equilibrium, i.e. when $N_-\ll N_+$. This means that the simple linear kinetic equation  is a result of  a non-linear dependence of the response ${\dot N}_-$ on the driving force  $X_-$.
In the considered case, one can calculate $\Xi(X_-)$ exactly. In the limit of large  $N_+$, the number of switches happening  over small time $\delta t$ is $N_+  \kappa \delta t$. 
Since all the switches are completely uncorrelated, the original formula for $\Sigma$--potential, Eq. (\ref{sigma}), results in the following expression for $\Xi(X)$: $\Xi= N_+\kappa\log [\cosh (2X_-)]$.
Here we have taken into account the fact that the change of population difference, $\delta N_-$ is either $2$ or $-2$ for any individual switch and the both possibilities are equally weighted.  Indeed,  any microscopic trajectory (sequence of individual switches) can be reversed, and this does not change the position of its midpoint, $\overline{N_-}\equiv (N_-(t_i)+N_-(t_f))/2$. After using Eq.(\ref{ham}),   we recover the expected linear equation for $N_-$.

We now proceed with the discussion of fluctuations around the  deterministic dynamics. Let ${\bf A^{(0)}}(t)$ be a solution to equation of motion (\ref{lagr}), and ${\bf a}(t)\equiv {\bf A}(t)-{\bf A^{(0)}}(t)$ is the deviation of an actual trajectory from it (${\bf a}(t_i)=0$). The corresponding deviation of the generalized entropy from its local maximum is given by the following quadratic expression:
\begin{eqnarray}
\label{gauss}
\delta S\left[{\bf a}(t)\right]=-\int_{t_i}^{t_f}    \frac{{\hat \lambda}^{-1}}{2}\cdot\left({\bf \dot a}+{\hat \lambda}\cdot{\hat \kappa}\cdot{\bf  a}\right)^2 dt =
\nonumber
\\
-\frac {1}{2}\int {\bf  a}_\omega \cdot\left(\omega^2{\hat \lambda}^{-1}+{\hat \kappa}\cdot {\hat \lambda}\cdot{\hat \kappa}\right)\cdot{\bf  a}_{-\omega}\frac{d\omega}{2\pi},
\end{eqnarray}
here  ${\hat \lambda} \equiv \delta^2 \Xi /\delta{\bf X}^2$ and ${\hat \kappa}\equiv \frac{1}{2}\delta^2 S/\delta {\bf A}^2$. Note that the deterministic trajectory is stable only if the response matrix ${\hat \lambda}\cdot{\hat \kappa}$ is positive-definite. Otherwise, any two trajectories originating from the same point diverge exponentially fast. 
By definition, $\exp \left(\delta S\left[{\bf a}(t)\right]\right)$ is the statistical weight of a given trajectory. Therefore, the  above quadratic functional, Eq. (\ref{gauss}), corresponds  to Gaussian statistics of  $\bf a$, with   $\langle{\bf  a}_{ \omega}{\bf  a}_{-\omega}\rangle=\left(\omega^2{\hat \lambda}^{-1}+{\hat \kappa}\cdot {\hat \lambda}\cdot{\hat \kappa}\right)^{-1}$. Equivalently, this result can be represented  by introduction  of a  random Gaussian noise ${\bf \eta}(t)$ to the second equation of (\ref{ham}), with  $\langle{\bf  \eta}(t){\bf  \eta}(t')\rangle={\hat \lambda}\delta(t-t')$. 

The fact that the same matrix $\delta^2 \Xi/\delta {\bf X}^2$ controls  both the strength of the fluctuations and the response to a small variation in driving force  (see (\ref{gauss}) or (\ref{ham})), is a signature of  FDT \cite{Groor}\cite{Chai}, which is conventionally applied  only in linear-response regime. In order to extend FDT to our far-from-equilibrium case, we probe the system with time-dependent perturbation introduced as an addition to entropy: $S({\bf A},t)=S_0({\bf A})+{\bf h}(t)\cdot{\bf A}$. In a particular case of a Hamiltonian system with fast coupling to a  thermal bath, $h$ is proportional to  field ${\bf h'}$ canonically-conjugated  to $\bf A$:   ${\bf h}=-\beta {\bf h'}$. The perturbation results in adding  a coupling term $\sum_\omega i\omega {\bf h}_{-\omega}{\bf a}_\omega/2$ to functional $\delta S$, Eq.(\ref{gauss}).
The response now can be determined by variation of $\delta S$ with respect to ${\bf a}_\omega$. An important aspect of this procedure is that the causality principle should be taken into account: in $t$-representation, ${\bf a}(t)$ should be varied with the specified past and unknown future (as we did while deriving equations of motion, (\ref{lagr}) and (\ref{ham})).   
In $\omega$ representation, the response is given by  
\beq
{\hat \Gamma} (\omega)\equiv \frac{\delta {\bf a}_\omega}{\delta {\bf h}_\omega}=\frac{i \omega}{2}\langle{\bf  a}_{ \omega}{\bf  a}_{-\omega}\rangle_{\cal  R}.
\eeq
Here index ${\cal  R}$ stands for the retarded part of the correlator, i.e. the one with all poles at  $\Im (\omega) >0$ half-plane. The above relationship extends the classical FDT towards strongly non-equilibrium regime. It has to be stressed  that ${\hat \Gamma} (\omega)$ does determine the response to small perturbations but, in contrast to linear response regime,  it {\em does not} relate the total  driving force $\delta S/\delta {\bf X}$ to the evolution rate $\bf \dot A$ (the relationships is given by {\em non-linear} Eq. (\ref{lagr}) or (\ref{ham})). 

Surprisingly enough, the applicability of FDT might be extended even further, to time scales comparable or shorter than ergodicity time $\tau_0$. Namely, if the frequencies of interest are considerably higher than the characteristic relaxation rates  (eigenvalues of relaxation matrix ${\hat \lambda}\cdot{\hat \kappa}$), we may neglect the dependency of the the driving force  on $\bf a$. In this case, the generalized entropy in $({\bf a},{\bf x})$--representation can be written in the following form:
\beq
\label{s_prime}
{\cal S}'=\sigma[{\bf x}(t)]-\sigma[{\bf h}(t)/2]+\int_{t_i}^{t_f}{\bf\dot a}\cdot\left(\frac{{\bf h}(t)}{2}-{\bf x}(t)\right)dt.
\eeq
Here ${\bf x}(t)$ is the deviation of ${\bf X}$ from its deterministic value ${\bf X}_0$, and  $\sigma[{\bf x}(t)]\equiv \Sigma[{\bf X}_0+{\bf x}(t)]-\Sigma[{\bf X}_0]$ is the corresponding deviation of  $\Sigma$-functional, which is no longer assumed to be local, i.e. $\omega\tau_0$ is not small.  In  this regime, to which one may  refer  as {\em sub-ergodic}, the statistics of $ {\bf  a}$ need not to be Gaussian. 

It  is easy to show that $\sigma[{\bf x}(t)]$ is the generating functional for ${\bf\dot a}$, i.e.  its variations coincide with the corresponding correlators of ${\bf\dot a}(t)$. On the other hand, in accordance with  (\ref{s_prime}), 
these variations of $\sigma$ determine the response of the system in {\em any order} of ${\bf h}$, ${\hat \Gamma}^{(n)} (\omega_1...\omega_n)\equiv\delta^n {\bf a}_{\Sigma\omega_k}/\delta {\bf h}_{\omega_1}...\delta {\bf h}_{\omega_n}$. This results in the following extension of FDT to sub-ergodic time scales:   
\beq
\label{FDT2}
{\hat \Gamma}^{(n)} (\omega_1...\omega_n) =\left\langle\left\langle{\bf  a}_{-\Sigma\omega_k}{\bf  a}_{\omega_1}...{\bf  a}_{\omega_n}\right\rangle\right\rangle_{\cal  R}\prod_{k=1}^n\frac{i \omega_k}{2}.
\eeq
This version of FDT is remarkable: (i) it may be applicable to systems with considerable memory effects, e.g. glasses; (ii) it establishes the relationship between the non-linear response of a  system and the deviation of its fluctuation statistics from Gaussian. The above  relationship is very similar to recent results for Markovian stochastic processes, \cite{Eyink2}. It should be noted that the direct experimental or numeric check of Eq. (\ref{FDT2}) may be difficult to perform without correct interpretation of field $\bf h$ (its physical meaning is straightforward only in the case of ideal (fast) thermal bath coupled to the system). In particular, such interpretation may involve frequency-dependent temperature\cite{Kurchan}.       

In conclusion, we have proposed a framework for  construction  of non-equilibrium macroscopic theory  of a complex system, starting with its fundamental non-dissipative dynamics.
This  approach   assumes  the  possibility of choosing a set of relatively slow (adiabatic) variables. Our major results  include the general form of equation of motion of the system under a given thermodynamic driving force (GLE) and extended FDT. Among the immediate possible applications of our scheme is the  development of non-equilibrium statistical theories of various complex systems starting with their model  Hamiltonians, such as   Heisenberg and $XY$ models \cite{XY}\cite{Chai}, or Gross-Pitaevsky model of Bose condensate.
Another intriguing direction of the development  of the generalized entropy approach is its use for Landau-type description of bifurcations. It also provides us with   an apparatus  to study the problem of kinetic tunneling between various steady states (attractors) of a non-equilibrium system.  A natural  extension  of our theory would be  its quantum generalization.      

{\bf Acknowledgement} The author is grateful to  T. Witten,  P. Cvitanovic, A. Sengupta, B. Shraiman, Y. Rabin,  C. Varma, E. Balkovski for useful discussions.

\end{document}